\documentclass[prb,twocolumn,superscriptaddress]{revtex4}
\usepackage{color}
\usepackage[dvips]{graphicx}
\begin{document}

\title{Singlet-triplet excitations and high-field magnetization in CuTe$_2$O$_5$}

\author {Zhe Wang}
\author{M.~Schmidt}
\affiliation{Experimental Physics V, Center for Electronic
Correlations and Magnetism, Institute for Physics, Augsburg
University, D-86135 Augsburg, Germany}
\author{Y. Goncharov}
\affiliation{Experimental Physics V, Center for Electronic
Correlations and Magnetism, Institute for Physics, Augsburg
University, D-86135 Augsburg, Germany} \affiliation{General Physics
Institute of the Russian Academy of Sciences, 119991 Moscow, Russia}
\author {Y. Skourski}
\author {J. Wosnitza}
\affiliation{Hochfeld-Magnetlabor Dresden (HLD), Helmholtz-Zentrum
Dresden-Rossendorf, D-01314 Dresden, Germany}
\author{H.~Berger}
\affiliation{Institute de Physique de la Mati\`{e}re Complexe, EPFL,
CH-1015 Lausanne, Switzerland}
\author {H.-A. Krug von Nidda}
\author {A. Loidl}
\author {J. Deisenhofer}
\affiliation{Experimental Physics V, Center for Electronic
Correlations and Magnetism, Institute for Physics, Augsburg
University, D-86135 Augsburg, Germany}

%\author{JPSJ Editorial Division, The Physical Society of Japan\thanks{E-mail address: jpsj{\_}edit@jps.or.jp}, and \name{Taro \surname{Butsuri}} %\\
% $^{1}Nihon Butsuri Gakkai}

%\inst{\address{2-31-22-5F Yushima, Bunkyo-ku, Tokyo 113-0034} %\\
%$^{1}$5-34-3-5F Shinbashi, Minato-ku, Tokyo 105-0004}

\begin{abstract}
By measuring the THz electron spin resonance (ESR) transmission spectra and high-field magnetization on the spin-gapped system CuTe$_2$O$_5$, we identified the singlet-triplet excitations in the dimerized non-magnetic ground state. The determined spin-gap value of $h\nu_0=4.94$~meV at the $\Gamma$ point ($\mathbf{Q}\simeq\mathbf{0}$) is significantly smaller than the strongest antiferromagnetic exchange interaction between the Cu ions predicted by theoretical investigations. We also observed the critical field $H_{c1}^{a^*}=37.6$~T for \textbf{H} $\bot$ \emph{bc}-plane and $H_{c1}^{bc}=40.6$~T for \textbf{H} $\|$ \emph{bc}-plane at the onset of non-zero magnetization, consistent with the gap value and corresponding anisotropic \emph{g}-factors determined previously. The observed singlet-triplet excitations in Faraday and Voigt configurations suggest a mixing of the singlet state with the $S_z=0$ triplet state and the $S_z=\pm 1$ triplet states, respectively, due to the Dzyaloshinskii-Moriya (DM) interaction with a DM vector perpendicular to the crystalline \emph{bc}-plane.

\end{abstract}

%\kword{spin dimer, magnetic excitations, CuTe$_2$O$_5$, THz transmission spectroscopy, electron spin resonance, high-field magnetization, Dzyaloshinskii-Moriya interaction}

\newcommand{\CuTeO}{CuTe$_2$O$_5$}

\maketitle

\section{Introduction}

During the past decades, transition-metal compounds based on Cu$^{2+}$ ions with a $3d^9$ configuration and spin-$1/2$ have been intensively studied due to the observation of various exotic collective phenomena in these systems.\cite{Lemmens03,Giamarchi08} A spin gap has been observed in those systems consisting of structural Cu dimers with predominant antiferromagnetic intradimer interaction, which is crucial for the occurrence of Bose-Einstein condensation of magnons such as in TlCuCl$_3$\cite{Nikuni00,Tanaka01} and BaCuSi$_2$O$_6$.\cite{Jaime04} In another type of spin-gapped systems based on Cu dimers, \CuTeO~as a particular example, the exchange interaction within the structural Cu dimer is not predominant, whereas the complex exchange paths between Cu ions mediated by the lone-pair cation of Te$^{4+}$ were found to be even stronger.\cite{Deisenhofer} Due to these complex exchange paths, it is still under debate whether the magnetic structure of \CuTeO~is an alternating spin-chain (i.e., one-dimensional) or an essentially two-dimensional coupled dimer system.\cite{Deisenhofer,Das,Eremina08,Gavrilova,Eremina11,Ushakov}

%such as spin-Peierls transition in CuGeO$_3$,\cite{Hase} magnetic-field-induced Bose-Einstein condensation of magnons in TlCuCl$_3$\cite{Nikuni00,Ruegg03} and Cs$_2$CuCl$_4$,\cite{Radu05} and cuprate superconductors.

%Figure 1
\begin{figure}[b]
\centering
\includegraphics[width=80mm,clip]{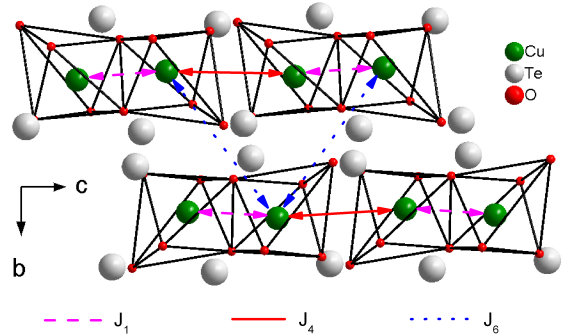}
\vspace{2mm} \caption[]{\label{Fig:CrysStruc} (Color online) Crystal structure of \CuTeO~ with space group $P2_1/c$. Structural dimers consisting of edge-sharing octahedra Cu$_2$O$_{10}$ are separated by Te ions. $J_1$, $J_4$, and $J_6$ are the exchange interactions between the first-, fourth-, and sixth-nearest-neighboring Cu ions, respectively.}
\end{figure}

%Figure 2
\begin{figure*}[t]
\centering
\includegraphics[width=160mm,clip]{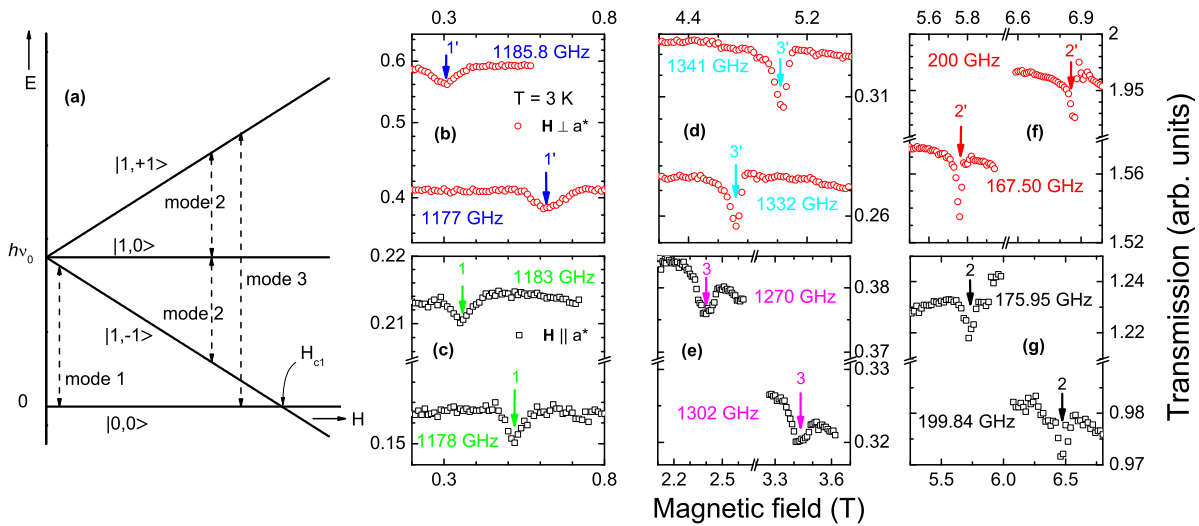}
\vspace{2mm} \caption[]{\label{Fig:EnergyScheme} (Color online) (a) Spin singlet $|0,0\rangle$ and Zeeman splitting of triplet states $|1,0\rangle$, $|1,-1\rangle$, and $|1,+1\rangle$ in a magnetic field \textbf{H}. Modes 1 and 3 are excitations from singlet to triplet states; Modes 2 are transitions between split-triplet states. $H_{c1}$ is explained in the text. (b)-(g) Transmission spectra measured at 3~K as a function of \textbf{H} at various frequencies corresponding to modes 1, 3, and 2 for \textbf{H} $\|$ $a^*$, and mode 1', 3', and 2' for \textbf{H} $\bot$ $a^*$.}
\end{figure*}

As shown in Fig.~\ref{Fig:CrysStruc}, \CuTeO~exhibits a monoclinic crystal structure with space group $P2_1/c$ and lattice parameters $a=6.817$~{\AA}, $b=9.322$~{\AA}, $c=7.602$~{\AA}, and $\beta=109.08^\circ$.\cite{Hanke} There are four Cu ions in one unit cell. Each Cu ion is surrounded by six oxygen atoms forming a strongly distorted octahedron. Two neighboring octahedra along the \emph{c} axis sharing an edge form structural dimer units Cu$_2$O$_{10}$, which are separated by Te-O bridging ligands. The high-temperature magnetic susceptibility of \CuTeO~can be fitted by a Curie-Weiss law with a Curie-Weiss temperature of $\Theta=-41$~K.\cite{Lemmens03} At $T=56.6$~K the susceptibility shows a maximum followed by a strong decrease to lower temperatures,\cite{Lemmens03} which is typical for magnetic dimer systems.\cite{B.B.} However, the susceptibility cannot be well fitted by the model of isolated magnetic dimers.\cite{Deisenhofer} Moreover, the isolated dimer model also faced difficulties in explaining the results of electron spin resonance (ESR). The extended H\"{u}ckel tight-binding (EHTB) method has been used to investigate the possible exchange paths.\cite{Deisenhofer} It has been shown that the exchange interaction between the sixth-nearest-neighboring Cu ions by a single O-Te-O bridge ($J_6$) is strongest and the nearest-neighboring interaction within the structural dimer ($J_1$) is the second strongest, while other exchange paths lead to much smaller exchange interactions between the Cu ions (see Fig.~\ref{Fig:CrysStruc} for exchange paths). In contrast to the results of EHTB, \emph{ab initio} density-functional theory (DFT) calculations within the local-density approximation (LDA) claimed that the strongest exchange interaction is between the fourth-nearest-neighboring Cu ions by two O-Te-O bridges ($J_4$, see Fig.~\ref{Fig:CrysStruc}) with the second and third strongest being $J_6$ and $J_1$, respectively,\cite{Das} indicating that \CuTeO~is a two-dimensional coupled spin-dimer system. Direct computation of exchange constants by an LDA+\emph{U} approach qualitatively showed a similar result,\cite{Ushakov} but determined a smaller value of the leading exchange constant $J_4=5.5$~meV ($\sim64$~K) than the value of $J_4=7.96$~meV ($\sim92.4$~K) obtained by quantum Monte Carlo (QMC) simulations.\cite{Das} In this paper, we report on the results of high-field ESR transmission spectroscopy up to $1.4$~THz and high-field magnetization measurements up to 60~T in \CuTeO. A spin singlet-to-triplet excitation of 4.94~meV has been resolved, which is considerably smaller than the strongest exchange interaction obtained by mean-field approximation\cite{Deisenhofer} or QMC.\cite{Das}

%You can use this file as a template to prepare your manuscript for \textit{Journal of Physical Society of Japan} (JPSJ)\cite{jpsj,instructions}. No sections or appendices should be given to other categories than Regular Papers. Key words are not necessary.

%Copy \verb|jpsj3.cls|, \verb|cite.sty|, and \verb|overcite.sty| onto an arbitrary directory under the texmf tree, for example, \verb|$texmf/tex/latex/jpsj|. If you have already obtained \verb|cite.sty| and \verb|overcite.sty|, you do not need to copy them.

%Many useful commands for equations are available because \verb|jpsj3.cls| automatically loads the \verb|amsmath| package. Please refer to reference books on \LaTeX\ for details on the \verb|amsmath| package.

%\begin{figure}[h]
%\centering
%\includegraphics[width=60mm,clip]{CuTe2O5_dimersnew.eps}
%\vspace{2mm} \caption{Projection of the crystal structure of
%CuTe$_2$O$_5$ on the $bc$-plane.}\label{fig:structure}
%\end{figure}

\section{Experimental Details}

Single crystals of CuTe$_2$O$_5$ were grown by a halogen vapor
transport technique, using HBr as transport agent. The charge and
growth-zone temperatures were 580~$^\circ$C and 450~$^\circ$C,
respectively. The stoichiometry of the single crystal was probed by
electron-probe microanalysis and X-ray diffraction. THz transmission experiments
were performed in the Faraday (propagating vector $\vec{k}$ $\|$ \textbf{H} $\bot$ \emph{bc}-plane) and Voigt configuration ($\vec{k}$ $\bot$ \textbf{H} $\|$ \emph{bc}-plane)
using a Mach-Zehnder-type interferometer
with backward-wave oscillators covering the frequency range 135~GHz-1.35~THz and a magneto-optical cryostat (Oxford/Spectromag) with
applied static magnetic field \textbf{H} up to 7~T. Magnetization measurements were performed in pulsed magnetic fields up to 60~T at the Dresden
High Magnetic Field Laboratory.\cite{Wosnitza07}

\section{Results and Discussion}

An isolated spin dimer with antiferromagnetic intradimer interaction exhibits a non-magnetic spin-singlet ground state and a three-fold degenerate spin-triplet excited state. In an applied magnetic field, the triplet state is completely split due to Zeeman interaction (see Fig.~\ref{Fig:EnergyScheme}(a)), thus there are three distinct types of excitations in a spin dimer. Modes 1 and 3 correspond to excitations between the singlet state $|0,0\rangle$ and the triplet states $|1,-1\rangle$, and $|1,+1\rangle$, respectively. Mode 2 corresponds to excitations between the split triplet states $|1,-1\rangle$ and $|1,0\rangle$, or between $|1,0\rangle$ and $|1,+1\rangle$.

%\begin{figure}[t]
%\centering
%\includegraphics[width=80mm,clip]{Zeeman_splitting.eps}
%\vspace{2mm} \caption[]{\label{Fig:EnergyScheme} (Color online)
%Zeeman-splitting of the triplet states in a magnetic field. Modes 1, 2 and 3 are described in the text.}
%\end{figure}

Figures \ref{Fig:EnergyScheme}(b)-(g) shows the ESR transmission spectra as a function of magnetic field measured at different frequencies with \textbf{H} $\|$ $a^*$ or \textbf{H} $\bot$ $a^*$, where $a^*$ denotes the direction perpendicular to the \emph{bc}-plane. Absorptions can be clearly observed in these spectra as marked by the arrows. The following features can be directly extracted from the data:

(i) \emph{Energy scale}. The absorptions shown in Fig.~\ref{Fig:EnergyScheme}(b)-(e) are obtained at the high-frequency range ($f>1$~THz) with relatively small field ($H<5.2$~T). In contrast, the absorptions in Fig.~\ref{Fig:EnergyScheme}(f) and (g) have much lower frequencies ($f\leq200$~GHz) and high fields ($H>5.6$~T). As suggested by the magnetic susceptibility and LDA+\emph{U} calculations,\cite{Lemmens03,Deisenhofer,Das,Ushakov} the leading antiferromagnetic interaction $J_4$ should not be smaller than 3.7~meV ($\sim 0.9$~THz), while the Zeeman energy is about 0.85~meV ($\sim 206$~GHz) at 7~T for a \emph{g}-factor of 2.1.\cite{Deisenhofer} Therefore, we can distinguish mode 2 in Fig.~\ref{Fig:EnergyScheme}(f) and (g) from modes 1 and 3 in Fig.~\ref{Fig:EnergyScheme}(b)-(e).

(ii) \emph{Field dependence}. The resonance field shifts to larger value with higher photon frequency in Fig.~\ref{Fig:EnergyScheme}(f) and (g) confirming the assignment of mode 2. A similar frequency dependence found in Fig.~\ref{Fig:EnergyScheme}(d) and (e) enables us to assign the mode 3, while the reverse situation in Fig.~\ref{Fig:EnergyScheme}(b) and (c) is a feature of mode 1 [see Fig.~\ref{Fig:EnergyScheme}(a)].

(iii) \emph{Orientation dependence}. The resonance field obtained at 200~GHz with \textbf{H} $\bot$ $a^*$ [Fig.~\ref{Fig:EnergyScheme}(f)] shifts strongly away from that measured at 199.84~GHz with \textbf{H} $\|$ $a^*$ [Fig.~\ref{Fig:EnergyScheme}(g)]. Since the frequencies are essentially the same, the large shift of resonance field with the variation of field orientation can be ascribed to the strong anisotropy of the \emph{g}-factor as explained in the following. A similar feature can be observed by comparing the lines measured with different orientations of applied field in Fig.~\ref{Fig:EnergyScheme}(b) and (c).

According to these features, the ESR modes can be unambiguously identified as mode 1, 2, or 3 for \textbf{H} $\|$ $a^*$, and mode 1', 2', or 3' for \textbf{H} $\bot$ $a^*$, as marked by the arrows in Fig.~\ref{Fig:EnergyScheme}(b)-(g).

The temperature-dependent behavior of the gapless intra-triplet modes has been reported previously.\cite{Deisenhofer} The intensity of the gapless mode decreases exponentially with decreasing temperature below 50~K. This is a typical feature of a spin-dimer system owing to the depopulation of the spin-triplet state at lower temperatures, which is also observed in the compound based on Cr$^{5+}$ dimers.\cite{Wang11} Accordingly, the intensity of the gapped mode increases towards lower temperature.

\begin{figure}[t]
\centering
\includegraphics[width=80mm,clip]{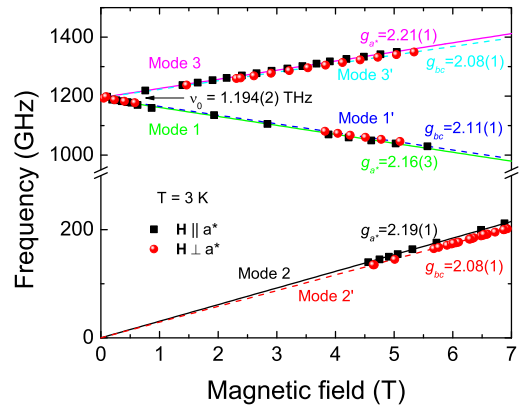}
\vspace{2mm} \caption[]{\label{Fig:SingletToTriplet} (Color online) Resonance frequencies of the observed absorption lines as a function of applied magnetic field for the field parallel and perpendicular to $a^*$.}
\end{figure}

The photon frequencies, at which the absorption lines were observed, are summarized as a function of the corresponding resonance fields in Fig.~\ref{Fig:SingletToTriplet} for \textbf{H} $\|$ $a^*$ and \textbf{H} $\bot$ $a^*$. The observed modes differ for the two orientations of magnetic field. As intra-triplet excitations, modes 2 and 2' are fitted by lines through the origin, resulting in the effective \emph{g}-factors $g_{a^*}=2.19(1)$ and $g_{bc}=2.08(1)$ for \textbf{H} $\|$ $a^*$ and \textbf{H} $\bot$ $a^*$, respectively. The \emph{g}-factors are consistent with the values $g_{a^*}=2.27(2)$, and $g_{bc}=2.11(3)$ obtained from \emph{X}- and \emph{Q}-band ESR, from which a nearly constant \emph{g}-factor in the \emph{bc}-plane was also demonstrated.\cite{Deisenhofer} Modes 1(1') and 3(3') can be described in terms of a linear Zeeman splitting following $h\nu =h\nu _{0}\mp g\mu_{B}H$ with $g_{a^*}=2.16\pm0.03$ ($g_{bc}=2.11\pm0.01$) and $g_{a^*}=2.21\pm0.01$ ($g_{bc}=2.08\pm0.01$), respectively, and a coincident zero-field value of $\nu_{0}=1.194(2)$~THz~($\sim4.94$~meV). These \emph{g}-factors, slightly larger than the spin-only value of $g=2$, are typical for the $3d^9$ electron configuration of Cu$^{2+}$ in distorted oxygen octahedra, where the orbital momentum is nearly quenched by the crystal field.\cite{Abragam1970} Both LDA and EHTB calculations have revealed that the important exchange paths in \CuTeO~are lying in the layers parallel to the \emph{bc}-plane, and the inter-layer exchange interactions between the Cu ions are much smaller and can be neglected. Therefore, we consider a Hamiltonian following the work by Leuenberger \emph{et al}.\cite{Leuenberger} $\mathcal{H}=\sum_{i}J_4\vec{S}_{i1}\cdot\vec{S}_{i2}+\sum_{(i,j')}J_6\vec{S}_{i1}\cdot\vec{S}_{j'2}+\sum_{[i,j]}J_1\vec{S}_{i1}\cdot\vec{S}_{j2}$, by taking into account only the three leading intra-layer interactions $J_4$, $J_6$, and $J_1$, where $i$ numerates the magnetic dimers (the subscripts \emph{1} and \emph{2} designate the two Cu ions in one dimer), $(i,j')$ denotes the pairs of neighboring magnetic dimers correlated by $J_6$, and $[i,j]$ counts the pairs of magnetic dimers correlated by $J_1$ (see Fig.~\ref{Fig:CrysStruc}). Using the standard-basis operator method within the random phase approximation,\cite{HaleyErdos} which has been applied to study the dispersion relation in several coupled magnetic dimer systems, such as Cs$_3$Cr$_2$Br$_9$,\cite{Leuenberger} BaCuSi$_2$O$_6$,\cite{Sasago97} Ba$_{3}$Cr$_2$O$_8$,\cite{Kofu09a} and Sr$_{3}$Cr$_2$O$_8$,\cite{Wang11,Castro10} the energy corresponding to the singlet-to-triplet excitation at the $\Gamma$ point ($\mathbf{Q}\simeq\mathbf{0}$) can be approximated by $h\nu_{0}\simeq\sqrt{J_{4}^{2}+J_{4}\gamma}$, where $\gamma=-2J_1+4J_6$. According to the LDA+\emph{U} calculations,\cite{Ushakov} the value of $h\nu_0=$5.32~meV determined for $U=10$~eV is close to the experimental result of 4.94~meV, while $h\nu_0=$7.83~meV is determined for $U=8$~eV which would underestimate the on-site Coulomb repulsion in \CuTeO. It is worth noting that only the first-order perturbation of $\gamma/J_4$ is considered here, although the ratio $J_6/J_4\sim0.3$ given by the LDA+\emph{U} calculation is not much smaller than 1.

%Figure 3

If there is only an isotropic intra-dimer exchange, the singlet-triplet excitation in a magnetic dimer cannot be observed in the ESR spectra. When a magnetic anisotropy via Dzyaloshiskii-Moriya (DM) interaction $\vec{D}\cdot(\vec{S_i}\times\vec{S_j})$ is present, the spin-triplet states are split even in zero field.\cite{Sakai} Moreover, the local triplet and singlet state will mix and allow the detectable optical singlet-triplet transitions.\cite{Room04} The singlet-triplet excitation modes 1 and 3 observed in the Faraday configuration (\textbf{H} $\|$ $a^*$) indicate a mixing of the singlet $|0,0\rangle$ and the triplet state $|1,0\rangle$ due to a DM vector $\vec{D}$ $\|$ \textbf{H} $\|$ $a^*$. For the Voigt configuration (\textbf{H} $\bot$ $a^*$), the DM interaction with $\vec{D}$ $\|$ $a^*$ $\bot$ \textbf{H} would mix $|0,0\rangle$ with $|1,\pm1\rangle$, thus the singlet-triplet excitations are also possible in this configuration, which complies with the observation of mode 1' and 3'.\cite{Room04,Matsumoto08} Therefore, we conclude that the DM vector is perpendicular to the crystalline \emph{bc}-plane.

%The anisotropic exchange interaction, i.e. $J^{xx}S_{i,x}S_{j,x}$,\cite{Eremina08} could also couple the local triplet state with singlet state, thus cannot be ruled out, although it is in the second order of the spin-orbital interaction.

\begin{figure}[t]
\centering
\includegraphics[width=80mm,clip]{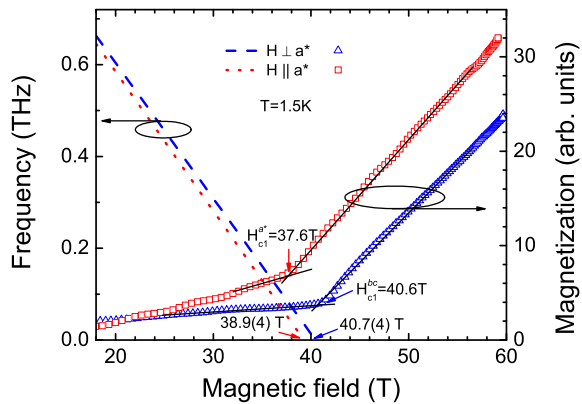}
\vspace{2mm} \caption[]{\label{Fig:Magnetization} (Color online) Left axis: Extrapolation of mode 1 (\textbf{H} $\|$ $a^*$, dotted line) and mode 1' (\textbf{H} $\bot$ $a^*$, dashed line) to the high-magnetic-field region. Right axis: Magnetization measured in a 60~T pulsed magnet at 1.5~K as a function of magnetic field for \textbf{H} $\|$ $a^*$ and for \textbf{H} $\bot$ $a^*$. Solid lines are guides for the eyes.}
\end{figure}

As illustrated in Fig.~\ref{Fig:EnergyScheme}(a), the Zeeman splitting will drive the lower-lying triplet state $|1,-1\rangle$ towards the singlet state and merge into the the ground state, when the magnetic field is sufficiently large and exceeds a critical field $H_{c1}= h\nu_0/g\mu_B$, with the Bohr magneton $\mu_B$ and the corresponding \emph{g}-factor. At $H=H_{c1}$, a transition from a non-magnetic ground state to one with a finite magnetization is induced by the field. The finite bulk magnetization should be observed and increase with increasing field $H>H_{c1}$ due to the larger population of the lower-lying triplet state. Figure \ref{Fig:Magnetization} shows the magnetization on the right ordinate as a function of magnetic field up to 60~T. A clear increase of magnetization can be observed at $H_{c1}^{a^*}=37.6$~T for \textbf{H} $\|$ $a^*$ and at $H_{c1}^{bc}=40.6$~T for \textbf{H} $\bot$ $a^*$, which agree with the values of 38.9(4)~T and 40.7(4)~T determined by $H_{c1}= h\nu_0/g\mu_B$ from mode 1 and 1' with $g_{a^*}$ and $g_{bc}$, respectively, shown on the left ordinate. This confirms the spin gap of $h\nu_0=4.94$~meV independently by means of high-field magnetization measurements.

%Figure 4

\section{Conclusions}
In summary, the magnetic properties of the spin-gapped system \CuTeO~have been studied by THz electron spin resonance transmission spectroscopy and high-field magnetization measurements. The excitations from singlet to Zeeman split-triplet states and the excitations between the split-triplet states have been observed, thus a spin-gap value of $h\nu_0=4.94$~meV is determined. A magnetic-field-induced transition from a non-magnetic to a finite magnetization state is observed at a critical field $H_{c1}^{a^*}=37.6$~T for \textbf{H} $\|$ $a^*$ and at $H_{c1}^{bc}=40.6$~T for \textbf{H} $\bot$ $a^*$ consistent with the anisotropic \emph{g}-factors and spin gap $h\nu_0$ determined from ESR spectra.

\begin{acknowledgments}
We thank M. V. Eremin of Kazan Federal University and Yuan Wan of Johns Hopkins University for fruitful discussions. We acknowledge the partial support by the DFG via TRR 80
(Augsburg-Munich) and FOR 960 (Quantum Phase Transitions). Part of this work has been supported by
EuroMagNET II under the EU contract 228043.
\end{acknowledgments}

\end{document}